\documentstyle{mn}

\input psfig

\newcommand{\gta}{\mathrel{\hbox{\rlap{\lower.55ex \hbox {$\sim$}}
                   \kern-.3em \raise.4ex \hbox{$>$}}}}
\newcommand{\lta}{\mathrel{\hbox{\rlap{\lower.55ex \hbox {$\sim$}}
                   \kern-.3em \raise.4ex \hbox{$<$}}}}

\newif\ifAMStwofonts

\ifoldfss
  \ifCUPmtlplainloaded \else
    \NewTextAlphabet{textbfit} {cmbxti10} {}
    \NewTextAlphabet{textbfss} {cmssbx10} {}
    \NewMathAlphabet{mathbfit} {cmbxti10} {} 
    \NewMathAlphabet{mathbfss} {cmssbx10} {} 
  \fi
  \ifAMStwofonts
    \ifCUPmtlplainloaded \else
      \NewSymbolFont{upmath} {eurm10}
      \NewSymbolFont{AMSa} {msam10}
      \NewMathSymbol{\upi}     {0}{upmath}{19}
      \NewMathSymbol{\umu}     {0}{upmath}{16}
      \NewMathSymbol{\upartial}{0}{upmath}{40}
      \NewMathSymbol{\leqslant}{3}{AMSa}{36}
      \NewMathSymbol{\geqslant}{3}{AMSa}{3E}

    \fi
  \fi
\fi 

\ifnfssone
  \newmathalphabet{\mathit}
  \addtoversion{normal}{\mathit}{cmr}{m}{it}
  \addtoversion{bold}{\mathit}{cmr}{bx}{it}
  \newmathalphabet{\mathbfit} 
  \addtoversion{normal}{\mathbfit}{cmr}{bx}{it}
  \addtoversion{bold}{\mathbfit}{cmr}{bx}{it}
  \newmathalphabet{\mathbfss} 
  \addtoversion{normal}{\mathbfss}{cmss}{bx}{n}
  \addtoversion{bold}{\mathbfss}{cmss}{bx}{n}
  \ifAMStwofonts
    \ifCUPmtlplainloaded \else
      \UseAMStwoboldmath
      \makeatletter
      \new@mathgroup\upmath@group
      \define@mathgroup\mv@normal\upmath@group{eur}{m}{n}
      \define@mathgroup\mv@bold\upmath@group{eur}{b}{n}
      \edef\UPM{\hexnumber\upmath@group}
      \new@mathgroup\amsa@group
      \define@mathgroup\mv@normal\amsa@group{msa}{m}{n}
      \define@mathgroup\mv@bold\amsa@group{msa}{m}{n}
      \edef\AMSa{\hexnumber\amsa@group}
      \makeatother
      \mathchardef\upi="0\UPM19
      \mathchardef\umu="0\UPM16
      \mathchardef\upartial="0\UPM40
      \mathchardef\leqslant="3\AMSa36
      \mathchardef\geqslant="3\AMSa3E
    \fi
  \fi
\fi 
\ifnfsstwo
  \DeclareMathAlphabet{\mathbfit}{OT1}{cmr}{bx}{it}
  \SetMathAlphabet\mathbfit{bold}{OT1}{cmr}{bx}{it}
  \DeclareMathAlphabet{\mathbfss}{OT1}{cmss}{bx}{n}
  \SetMathAlphabet\mathbfss{bold}{OT1}{cmss}{bx}{n}
  \ifAMStwofonts
    \ifCUPmtlplainloaded \else
      \DeclareSymbolFont{UPM}{U}{eur}{m}{n}
      \SetSymbolFont{UPM}{bold}{U}{eur}{b}{n}
      \DeclareSymbolFont{AMSa}{U}{msa}{m}{n}
      \DeclareMathSymbol{\upi}{0}{UPM}{"19}
      \DeclareMathSymbol{\umu}{0}{UPM}{"16}
      \DeclareMathSymbol{\upartial}{0}{UPM}{"40}
      \DeclareMathSymbol{\leqslant}{3}{AMSa}{"36}
      \DeclareMathSymbol{\geqslant}{3}{AMSa}{"3E}
    \fi
  \fi
\fi 

\ifCUPmtlplainloaded \else
  \ifAMStwofonts \else 
    \def\upi{\pi}
    \def\umu{\mu}
    \def\upartial{\partial}
  \fi
\fi

\def\aaa#1{{A\&A,} {#1}}

\def\apj#1{{ApJ,} {#1}}

\def\mnras#1{{MNRAS,} {#1}}

\newbox\grsign \setbox\grsign=\hbox{$>$}
\newdimen\grdimen \grdimen=\ht\grsign
\newbox\laxbox \newbox\gaxbox
\setbox\gaxbox=\hbox{\raise.5ex\hbox{$>$}\llap
     {\lower.5ex\hbox{$\sim$}}}\ht1=\grdimen\dp1=0pt
\setbox\laxbox=\hbox{\raise.5ex\hbox{$<$}\llap
     {\lower.5ex\hbox{$\sim$}}}\ht2=\grdimen\dp2=0pt
\def\gax{\mathrel{\copy\gaxbox}}

\def\lax{\mathrel{\copy\laxbox}}
\def\lta{\lax}
\def\gta{\gax}

\title[Unipolar outflows in accretion discs]
  {Unipolar outflows and global meridional 
   circulations in rotating accretion flows}

\author[I. V. Igumenshchev]
  {Igor V. Igumenshchev\thanks{E-mail: ivi@fy.chalmers.se}
     \thanks{On leave from: Institute of Astronomy, 
     48 Pyatnitskaya street, Moscow, 109017, Russia}\\
  Institute for Theoretical Physics, G\"oteborg University and 
  Chalmers University    
     of Technology, 412 96 G\"oteborg, Sweden
  }

\pagerange{\pageref{firstpage}--\pageref{lastpage}}
\pubyear{2000}

\begin{document}

\maketitle

\label{firstpage}

\begin{abstract}

Using two-dimensional simulations
of non-radiative viscous rotating black hole accretion flows,
we show that the flows with $\alpha\sim 0.1-0.3$
self-organize to form
stationary  unipolar or bipolar  outflows accompanied by
global meridional circulations. 
The required energy comes, with efficiency $\sim 0.001-0.01$, 
from the matter directly accreted onto black hole.
Observational implications are discussed.

\end{abstract}


\section{Introduction}

The considerable recent interest in models of geometrically thick 
rotating accretion flows has been motivated by ability of these models
to explain unusual observational properties of some accreting 
black hole candidates.
There are two major classes of such  models.
(a) At low accretion rates, the optically thin accretion flow is not able
to radiate efficiently and the internal energy stored in the flow
is advected into the black hole. Models of such flows have been developed 
by Ichimaru (1977), Rees et al. (1982), Narayan \& Yi (1994), 
Abramowicz et al. (1995), and others see in
recent reviews by Narayan, Mahadevan \& Quataert (1998) and
Kato, Fukue \& Mineshige (1998).
These models are called advection dominated accretion flows (ADAFs)
and have been applied for explanation of spectral
properties of low luminosity high-energy sources.
(b) At high accretion rates
the flow is optically thick. The liberated
binding energy is converted to radiation which is trapped by the flow
and advected into the black hole as shown by Katz (1977) and
Begelman (1978) for spherical accretion. 
Rotating accretion flows of this type are called thick discs,
Polish doughnuts, or slim discs. They have been developed by
Abramowicz, Jaroszy\'nski \& Sikora (1978),
Jaroszy\'nski, Abramowicz \& Paczy\'nski (1980), Abramowicz et al. (1988)
and others.

Most of the ADAF models have been constructed in a one-dimensional
approach which restricts properties of the solutions
by considering only the vertically-averaged purely inward motion:
the multidimensional character of flow is missing.
Narayan \& Yi (1995a) pointed out the possible importance
of polar outflows in ADAFs.
Analytic study of accretion flows
with polar outflows was performed by Xu \& Chen (1997) and
Blandford \& Begelman (1999) in a self-similar approach, and
numerical two-dimensional studies have
recently been carried out by 
Igumenshchev, Chen \& Abramowicz (1996),
Igumenshchev \& Abramowicz (1999, hereafter IA99) and 
Stone, Pringle \& Begelman (1999).
These numerical studies demonstrated that 
in the case of small or moderate viscosity ($\alpha\la 0.1$), non-radiative
accretion flows are convectively unstable and accompanied by irregular
bipolar outflows. They do not form powerful unbound winds.
At large viscosity ($0.1\la\alpha < 1$) flows are stable
and may have (but do not have to have) strong outflows.

In this Letter
we present results from two-dimension axisymmetric hydrodynamical simulations
of non-radiative rotating accretion flows with large viscosity. 
The flows form powerful unipolar or bipolar outflows and could be
stationary or nonstationary, depending both on $\alpha$ and 
on the adiabatic index
$\gamma$, and self-organize into global meridional circulation cell(s).
The energy required to support the circulation
is extracted from matter accreted into the black hole with
efficiency $\sim 10^{-3}-10^{-2}$. 

\section{Numerical method}

We compute models by solving the 
non-relativistic hydrodynamical equations
\begin{equation}
 {d \rho\over dt}+\rho\nabla\vec{v}=0,
\end{equation}
\begin{equation}
 \rho{d \vec{v}\over dt}=-\nabla P + \rho\nabla\Phi + \nabla{\bf \Pi},
\end{equation}
\begin{equation}
 \rho{d e \over dt}=-P\nabla\vec{v}+Q,
\end{equation}
where $\rho$ is the density, $\vec{v}$ is the velocity,
$\Phi=-GM/r$ is the Newtonian gravitational potential for a central point
mass $M$,
$e$ is the specific internal energy, ${\bf \Pi}$ is the 
viscous stress tensor with all components included,
and $Q$ is the dissipation function.
We adopt the ideal gas equation of state, $P=(\gamma-1)\rho e $,
and consider only the shear viscosity, assuming the 
$\alpha$-prescription,
\begin{equation}
 \nu=\alpha{c_s^2\over \Omega_K},
\end{equation}
where
$c_s=\sqrt{P/\rho}$ is the isothermal sound speed, and
$\Omega_K=\sqrt{GM/r^3}$ is the Keplerian angular velocity.
Details of the numerical technique used to solve (1)-(3) 
in axial symmetry were discussed by IA99.

We use a spherical grid 
$N_r\times N_\theta=130\times 50$ with the inner radius at $r_{in}=3 r_g$ and
the outer radius at $r_{out}=8000 r_g$, where $r_g=2GM/c^2$ is the 
gravitational radius of black hole.
The grid extends from $0$ to $\pi$ in the polar direction.
To model the relativistic Roche lobe overflow, that governs the flow
near to the black hole (Abramowicz 1981), in Newtonian gravity
we adopt absorbing 
boundary conditions at $r=r_{in}$ 
together with the condition of the derivatives
$d(v_\theta/r)/dr$ and $d(v_\phi/r)/dr$ being zero.
The latter means that there is no viscous energy flux
from the inner boundary associated with the ($r\theta$) and
($r\phi$) components of the shear stress. 
At $r_{out}$ we apply free outflow boundary conditions
by interpolating all dynamical variables behind $r_{out}$.

In the calculations we assume a source of matter with a constant ejection rate.
The source  is located around $\theta=\pi/2$ 
in the vicinity of $r_{out}$.
Matter is ejected there with angular momentum equal to
$0.95$ times the Keplerian angular momentum.
Due to the viscous spread, part of the ejected matter moves inwards
and forms the accretion flow. The other part leaves
the computation domain freely through the outer boundary.
We start the computation of a model from an initial state,
the choice of which is not crucial,
and follow the evolution until 
a quasi-stationary flow pattern is established.
Typically, this takes a few viscous time scales
estimated at $r_{out}$.

\section{Two-dimension hydrodynamical models}

\begin{table}
\caption[ ]{Parameters of two-dimension models.}
\begin{tabular}{cccccc}
\hline
 Model~~ &~~~$\alpha$~~~&~~~$\gamma$~~~&~~~outflow~~~~&~~~stability~~~&
~~~~$\epsilon$~~~~ \\
\hline
$A$ & $0.3$    & $5/3$ &  bipolar  &  stable  &  0.002  \\
$B$ & $0.1$    & $5/3$ &  unipolar &  stable  &  0.012  \\
$C$ & $0.3$    & $4/3$ &  unipolar &  quasi-stable  &  0.002  \\
$D$ & $0.1$    & $4/3$ &  bipolar  &  unstable      &  0.006  \\
\hline
\end{tabular}
\end{table}

Four models with various values of $\alpha$ and $\gamma$ are listed in Table~1,
which also lists the
type of the outflow, the stability and the efficiency $\epsilon$
defined by (8).
All of the models have 
powerful outflows, launched 
at radial distances $10-100 r_g$. Models~A and B are stationary,
and Model~C shows a stable time-averaged global flow pattern,
perturbed from time to time by 
hot convective bubbles that originate very close to $r_{in}$.
Model~D is less stationary, with significant convection activity
which originates in the innermost part of the flow.

\begin{figure}
\hbox{\psfig{file=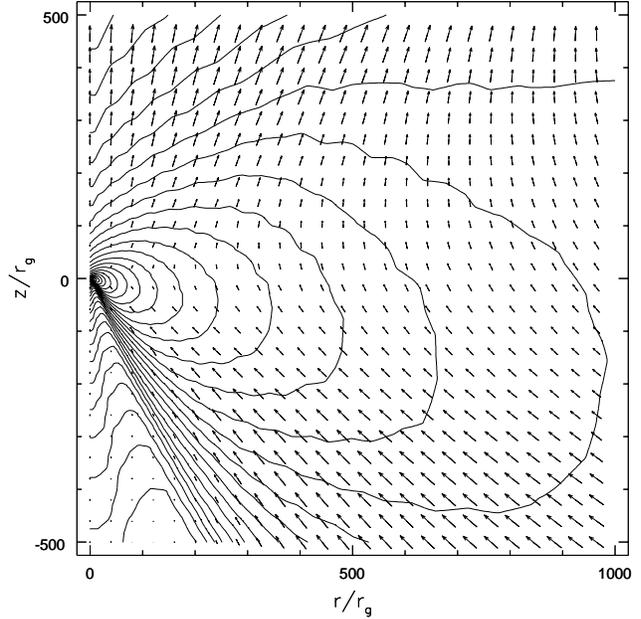,height=8.2cm}}
\caption{The flow pattern in Model~B. Distributions of density $\rho$ and
the vector $r^2\rho\vec{v}$ are shown in the meridional cross section.
The vertical axis coincides with the axis of rotation.
The black hole is located in the origin (0,0). 
The contour lines are spaced with $\Delta\log\rho=0.1$.
Unipolar outflow is clearly seen in the vectors. 
Only a small fraction of the matter
circulating in the computation domain directly accretes into the black hole.
}
\label{fig1}
\end{figure}

The flow patterns with bipolar outflows in Models~A and D are quite similar
to those discussed by IA99 (their Models~1 and 5, respectively),
despite differences in $\alpha$ and $\gamma$. 
The most interesting feature of Models~B and C is the unipolar
outflow. 
Figure~1 presents the flow pattern for Model~B.
Matter contained within the calculation domain
forms a global meridional circulation cell 
with the spatial scale $\sim r_{out}$.
Most of the stream-lines of the flow start at the source of matter at $r_{out}$,
go inward, approach some minimum radius near to the equatorial plane, 
turn outward,
and leave the computational domain through the upper hemisphere.
In Model~B we observe one large circulation cell 
(toroidal in three-dimensions), whereas in Model~C
the large circulation cell co-exists with a smaller one
of opposite vorticity.
The circulation is powered by the one-sided outflow generated in
the vicinity of black hole.
The part of the outflow which is close to the polar axis
is most efficiently accelerated and
becomes supersonic at a radial distance $\approx 1000 r_g$.
This part of the flow contains a small mass fraction and has
outward directed velocities, which are larger 
than the escape velocity, $v_r > v_{esc}=\sqrt{2GM/r}$. 
Obviously, it can escape to
infinity, even if cooling processes become efficient at large
distances $r\ga 1000 r_g$. 
However, most of the outflowing matter forms a `subsonic wind'.
The evolution
of this wind at large distances will be governed by cooling processes,
which are not considered in our models.

What drives such powerful outflows and circulations?
Obviously, the power is extracted from the rotating accretion flow
with the help of
a mechanism which redistributes energy and momentum
between different parts of fluid.
In our non-radiative viscous
models only the shear stress can provide such a mechanism.
Due to the stress, the inner more rapidly rotating parts of
the accretion flow pass a fraction of kinetic (not only rotational)
energy to the outer parts.

The importance of outward energy transport supported by convection
in geometrically thick accretion discs has been independently recognized 
by Bardeen (1973), Abramowicz (1974) and Bisnovatyi-Kogan \& Blinnikov (1977), 
and first studied in detail by
Paczy\'nski \& Abramowicz (1982) with a follow-up by
R\'o\.zyczka \& Muchotrzeb (1982). However, these models do not include
inward advection.
In later works 
(Begelman \& Meier 1982; Narayan \& Yi 1995a;
Honma 1996; Kato \& Nakamura 1998; Manmoto et al. 2000) advection
was included, but
the convective outward energy flux was found to be weak,
and was always dominated by the assumed purely
inward directed advective energy flux.
In the `self-similar' models of ADAFs (Narayan \& Yi 1994)
the viscous energy flux,
\begin{equation}
\dot{E}_{visc}(r)=-2\pi r^2\int_0^\pi\left(
  v_r\Pi_{rr}+v_\theta\Pi_{r \theta}
  +v_\phi\Pi_{r \phi}\right)\cos\theta d\theta,
\end{equation}
is positive, directed outward, and
is exactly balanced by the inward advection of energy with the corresponding
flux
\begin{equation}
\dot{E}_{adv}(r)=2\pi r^2\int_0^\pi\rho v_r\left({v^2\over 2}+
  W-{GM\over r}\right)\cos\theta d\theta,
\end{equation}
where $W$ is the specific enthalpy. Thus, the total energy flux,
\begin{equation}
\dot{E}=\dot{E}_{adv}+\dot{E}_{visc},
\end{equation}
is zero everywhere, $\dot{E}(r) = 0$.
Abramowicz, Lasota \& Igumenshchev (1999) demonstrated
that $\dot{E}=0$ is an artifact of the assumed self-similarity: 
ADAFs that obey physically reasonable boundary conditions
have, in general, $\dot{E} \neq 0$. 

Our models 
have a specific geometry of flow which is very different from pure 
equatorial inflow. 
In Figure~2 we show the flow pattern for Model~B in the vicinity
of the inner boundary.
In this figure the arrows show the velocity directions, and
the ellipse is the projection of the radius $r=R_A$ at the equatorial plane.
Matter, which crosses the equatorial plane inside $R_A$,
accretes into the black hole at the rate $\dot{M}$. 
Using the analogy of the flow pattern in Figure~2 with Bondi-Hoyle
accretion, we call $R_A$ the `accretion radius'.
In the case of Model~B, $R_A\approx 10 r_g$. Model~C has a similar flow
pattern, but $R_A$ varies with time around the average value $\approx 80 r_g$.

\begin{figure}
\hbox{\psfig{file=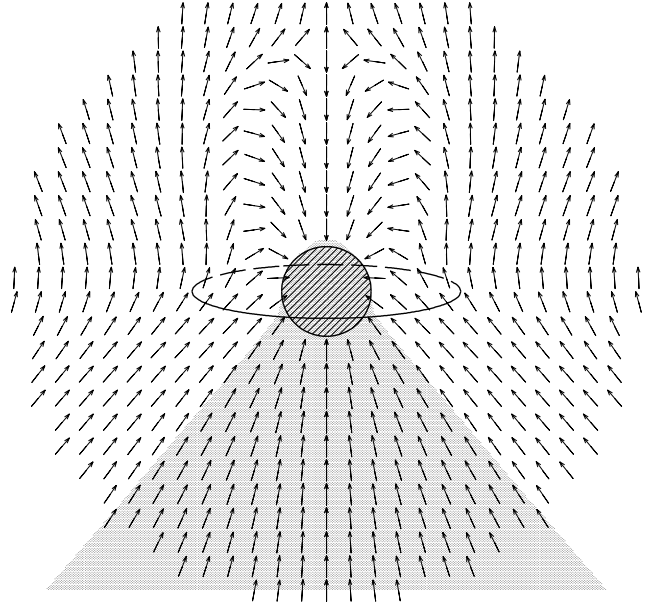,height=8.0cm}}
\caption{
Two-dimensional structure of accretion flow in the innermost part of Model~B.
The region with radius $20 r_g$ is shown.
The axis of symmetry of the picture coincides with the axis of rotation.
The central shadow circle indicates the location 
of the inner numerical boundary.
Arrows point to directions of motion.
The ellipse is the projection of a circle of `accretion' radius $R_A$  
at the equatorial plane.
Matter, which crosses the equatorial
plane inside $R_A$ is absorbed by black hole.
Regions with supersonic inward velocities are shown in grey.
}
\label{fig2}
\end{figure}

The flow geometry shown in Figure~2 allows a significant transport
of energy by shear stress from the matter accreted into the black hole 
to the matter
outflowing to the upper hemisphere. Indeed, the stream lines
of matter which is eventually inflowing and outflowing are located close
together until the accreting matter reaches $r\sim R_A$. During this phase,
efficient momentum and energy exchange takes place.
We shall characterize the outward energy transport $\dot{E}$
by the `accretion efficiency' 
\begin{equation}
\epsilon=\dot{E}/\dot{M}c^2,
\end{equation}
which is different from the 
standard definition of radiative efficiency,
\begin{equation}
\epsilon_{rad}=L/\dot{M}c^2,
\end{equation}
where $L$ is the total luminosity of accretion flow.

Assuming that matter falls freely into the black hole inside $R_A$
and that only the binding energy at the orbit $r=R_A$ is liberated,
one can obtain an estimate of the maximum accretion efficiency,
\begin{equation}
\epsilon\simeq {1\over 4}{r_g\over R_A}.
\end{equation}
In the case of Models~B and C, formula (10) predicts $\epsilon\approx 0.02$ 
and $0.004$, respectively. Both values are
significantly larger than the prediction of $\epsilon_{rad}$ 
for ADAFs,
$\epsilon_{rad}\la 10^{-4}$ (Narayan \& Yi 1995b).

\begin{figure}
\hbox{\psfig{file=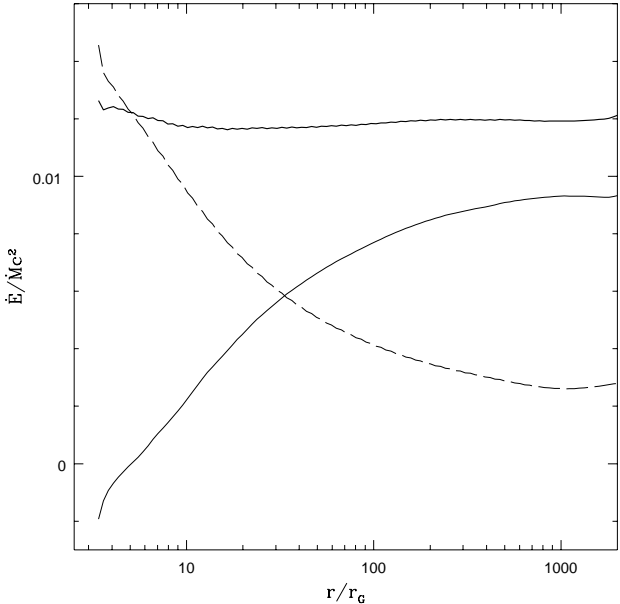,height=8.0cm}}
\caption{
Radial dependence of the total energy flux $\dot{E}$ (heavy line),
the advective energy flux $\dot{E}_{adv}$ (solid line) and
the viscous energy flux $\dot{E}_{visc}$ (dashed line) in Model~B.
The energy fluxes are defined by eqs. (7), (6) and (5), respectively.
}
\label{fig3}
\end{figure}

$\dot{E}(r)$, $\dot{E}_{adv}(r)$ and
$\dot{E}_{visc}(r)$ for Model~B are shown in Figure~3.
In a steady state, $\dot{E}$ must be constant.
The variation in $\dot{E}$ seen in Figure~3 is about $10\%$.
This can be partially explained by a small non-stationarity of the flow,
and small inaccuracies in our numerical code,
which does not exactly conserve the total energy.
Test calculations have shown that this inaccuracy 
gives an error of less than $5\%$.
From Figure~3, one can see that for Model~B,
$\epsilon\approx 0.01$, which 
is only a factor of 2 smaller than what is predicted by estimate (10).
The viscous flux $\dot{E}_{visc}$ (the dashed line in Fig.3) is always positive
and dominates the
advection flux $\dot{E}_{adv}$ (solid line) inside the radius $\approx 30 r_g$.
At larger radii the energy is transported outward mainly
by advection. Note, that $\dot{E}_{adv}$ 
changes sign at $r\approx 5 r_g$. Inside this radius 
the inward energy advection compensates a fraction of the outward viscous energy
flux.

In all models, the time averaged net accretion rate $\dot{M}(r)$ 
through successive spheres of radius $r$, is a constant
inside $\sim 1000 r_g$  to good accuracy.
However, the rates of mass inflow  $\dot{M}_{in}$ and outflow
$\dot{M}_{out}=\dot{M}_{in}-\dot{M}$ are both increasing functions
of $r$. In Model~B $\dot{M}_{in}(r)$ is well approximated 
by a power law 
with index $\beta\approx 0.5$ in the radial range $10-1000 r_g$.
Models~A, C and D show a similar power law behaviour for $\dot{M}_{in}(r)$,
but with slightly different indices.
Such a fast increase outward of $\dot{M}_{in}$ 
and $\dot{M}_{out}$
indicates the importance of global
circulation motions in the dynamics of accretion flow in the models considered. 
Only a small part of the 
matter circulated in the computational domain is accreted by the black hole.
Most of the matter in the case of (quasi) stable Models~A, B and C,
which starts at the source near to the
outer boundary, has escaped outside the outer boundary after one cycle of
circulation. This behaviour is in agreement
with the results of Blandford \& Begelman (1999), but unlike these authors,
we see powerful outflows only 
in the case of large viscosity, $\alpha\ga 0.1$.
We also note that the radial dependence of $\dot{M}_{in}$
and $\dot{M}_{out}$ in our simulations is in good agreement with
the results of Stone et al. (1999), although in thier models
accretion flows are dominated by small-scale vortices 
and circulation motions.

\section{Discussion}

Our numerical models can be applied to the
two well-studied accretion regimes mentioned in the Introduction:
optically thick and optically thin.
The matter in optically thick flows is radiation dominated and characterized by
$\gamma=4/3$, so Models~C and D can be relevant in this case. 
In optically thin flows the `effective' adiabatic index ranges
from about 3/2 to 5/3, depending on the strength of the magnetic field 
(see Narayan \& Yi 1995b).

Flow patterns of `optically thin' models consist of large-scale 
stable subsonic circulation cell(s) (in the $r-\theta$ plane):
two equatorially symmetric cells in Model~A and one cell in Model~B.
Only a small fraction of the outflowing matter forms the unbound supersonic
unipolar outflow in Model~B.
A fraction $\epsilon\sim 0.01$ 
of the total energy of the matter accreted by the black hole is extracted
and remains in the form of kinetic and thermal energy of the circulated matter.
Although the present numerical models include no energy losses,
one can speculate about the fate of subsonic outflows when 
energy losses are important, using the recent results obtained by
Allen, Di Matteo \& Fabian (1999) and Di Matteo et al. (1999).
They have noted that in optically thin
accretion flows coupled with strong outflows, the radiative energy losses
are dominated by bremsstrahlung.
The most efficient cooling in such accretion flows occurs at the maximum
radius where this flow exists. 
Taking into account this result and the results of our numerical simulations
we propose the following scenario of black hole accretion accompanied
by global meridional circulation of matter. 
Subsonic outflows originate near to the black hole and are mainly supported by
the pressure gradient force while the matter radiates weakly.
At large radial distances, where the dynamical time scale of flows becomes
comparable with the bremsstrahlung cooling time, the matter cools down and
its outward motion cannot be supported any longer by the pressure gradient.
The matter turns back and forms the inflowing part of the circulation cell(s).
A schematic illustration of such a flow pattern in the case of
one circulation cell near to the black hole is shown in Figure~4.
Assuming that all of the thermal energy 
carried by outflows is radiated away,
we can roughly estimate the radiative efficiency to be
$\epsilon_{rad}\approx\epsilon\sim 0.01$.
Note that this scenario can only be correct if the thermal instability
develops slowly at the outer region of circulation cell(s),
but this is what one would expect.

\begin{figure}
\hbox{\psfig{file=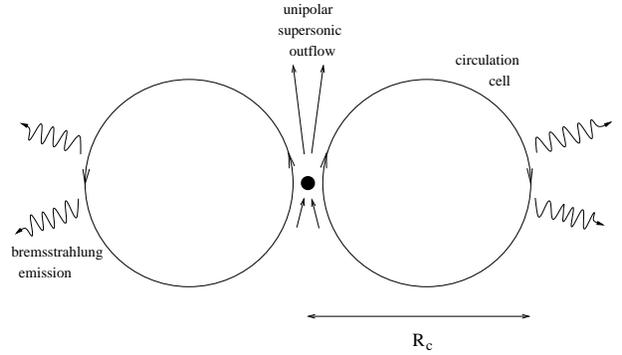,width=8.0cm}}
\caption{Schematic illustration of the proposed geometry of 
an optically thin rotating accretion flow
accompanied by the global meridional circulation of matter.
}
\label{fig4}
\end{figure}

Basic properties of accretion flows with circulations
can roughly be estimated.
Let $R_c$ be the spatial scale of the circulation cells which contain
matter with the average density $\rho_c$. 
Comparing the bremsstrahlung cooling time for 
matter with the virial temperature and 
the dynamical time scale $t_d=R_c/V_c$ (where $V_c=\beta V_K$ 
is the mean circulation
velocity at $R_c$, $V_K$ is the
Keplerian velocity and the value of the parameter 
$\beta\sim\alpha\sim 0.1$ is taken from numerical models), we can estimate
$\rho_c\propto R_c^{-2}$ and consequently, the mass involved in 
circulations,
\begin{equation}
M_c \sim \left({\beta\over 0.1}\right)
\left({R_c\over 10^3 r_g}\right)
\left({M\over 10^9 M_\odot}\right)^2 M_\odot,
\end{equation}
and the bremsstrahlung luminosity, 
\begin{equation}
{L\over L_{Edd}}\sim 3\cdot 10^{-5}\left({\beta\over 0.1}\right)^2 
\left({10^3 r_g\over R_c}\right)^{3/2}.
\end{equation}
Thus, the less massive and more compact object the higher luminosity
it has.
In a steady state, the external mass supply to the circulation cells 
is equal to
the mass accretion rate $\dot{M}=L/\epsilon c^2$.
Without the mass supply, the characteristic lifetime of the
circulation cells is
\begin{equation}
t_c={M_c\over\dot{M}}\sim 10^3\left({0.1\over\beta}\right)
\left({\epsilon\over 0.01}\right)\left({R_c\over 10^3 r_g}\right)^{5/2}
\left({M\over 10^9 M_\odot}\right) {\rm yrs.}
\end{equation}
Note, that in estimates (11)-(13), we ignore the mass and energy losses 
due to the 
supersonic unipolar/bipolar outflows, and due to a wind which
starts on the `outer surface' of circulation cells and
carries out all of the excess angular momentum. 

Estimates (11) and (12) agree quite well with the observed data for the core
of the elliptical galaxy M87 (Allen et al. 1999). Also, our finding of
unipolar outflows from accreting black holes can be used to explain
the one-sided jets observed in M87 and other objects.


\subsection*{Acknowledgments} We thank Marek Abramowicz and John Miller
for stimulating discussions and comments on drafts of this paper.
We gratefully acknowledge hospitality at the
International School for Advanced Studies in Trieste,
where a part of this work was done.

\label{lastpage}

\end{document}